\documentclass[angeo,ms]{copernicus}
\newcommand{\aap}{    {\it Astron. \& Astrophys.}}

\newcommand{\apj}{    {\it Astrophys. J.}}

\newcommand{\jgr}{    {\it J. Geophys. Res.}}

\newcommand{\solphys}{{\it Solar Phys.}}

\begin{document}

\title{Estimation of width and inclination of a filament sheet using He II 304 \AA\ observations by STEREO/EUVI}

\author[1]{Sanjay Gosain}
\author[2]{Brigitte Schmieder}

\affil[1]{Udaipur Solar Observatory, P. Box 198, Dewali, Udaipur 313001}
\affil[2]{Observatoire de Paris, LESIA, 92190 Meudon, France}

%% The [] brackets identify the author to the corresponding affiliation, 1, 2, 3, etc. should be inserted.

\runningtitle{Estimation of Filament Width}

\runningauthor{S. Gosain and B. Schmieder}

\correspondence{Sanjay Gosain\\ (sgosain@prl.res.in)}

\received{}
\pubdiscuss{} %% only important for two-stage journals
\revised{}
\accepted{}
\published{}

%% These dates will be inserted by the Publication Production Office during the typesetting process.

\firstpage{1}

\maketitle
%% only used for copernicus2.cls

\begin{abstract}
 The STEREO mission has been providing stereoscopic view of the filament eruptions in EUV wavelengths. The most extended view during
  filament eruptions is seen in He II 304 \AA\ observations, as the filament spine appears darker and sharper. The  projected
  filament width appears differently when viewed from different angles by STEREO satellites. Here, we present a method for estimating
  the width and inclination of the filament sheet using He II 304 \AA\ observations by STEREO-A and B satellites from the two viewpoints.
   The width of the filament sheet, when measured from its feet to its apex, gives estimate of filament height above the chromosphere.
\end{abstract}

\introduction
Filaments are vertical slabs or sheet-like plasma structures in the
corona. Their typical density ($\sim 2 \times 10^{-10} kg/m^3$) is
about 200 times higher and their temperature ($\sim$7,000 K) about 200 times
lower than the surrounding corona. The topology of the magnetic
field in filaments is such that the material is held in
equilibrium against gravity and is thermally insulated from the
surroundings. There are two types of filaments, one associated with
active regions and other associated with large-scale weak magnetic regions or located between them.
 The lifetime of the former is shorter, typically few hours
to days, while the latter are quite stable lasting for days to
several weeks and are called quiescent filaments. The active region
filaments are denser and their heights typically reach up to 10 Mm
 \citep{Schmieder1988}, while the quiescent filaments are diffuse or
less dense and reach greater heights of up to 200 Mm
\citep{Pettit1932,Azambuja1945,Schmieder2004,Schmieder2008}.
 However, instabilities in the filament equilibrium
sometimes lead to its eruption forming the spectacular {\it
disparition brusque} (DB) \citep{Forbes1991}. Most of the DBs
are associated with the CMEs  \citep{Gopalswamy2003}.
\cite{Mouradian1989} classified DBs into two categories, one
due to dynamic and the other due to thermal instability and
suggested that the DBs associated with CMEs are due to dynamic
instability. In case of filament disappearance due to thermal
instability the filament reappears after the plasma has cooled
down.
 The determination of true filament height is an important parameter in
studies related to filament eruption  in order to determine the
height-time profile during rapid-acceleration phase of filament
eruption \citep{Schrijver2008}. The stability of
prominences/filaments depends upon their so-called critical
height, above which they become unstable and erupt
\citep{Filippov2001}. In order to verify the critical-height
criteria for erupting prominence one needs ways of determining
filament height even while filament is on-disk.

Earlier, the determination of filament height was possible only
from the observations of filament at limb i.e., as a
prominence. With the advent of STEREO mission
\citep{Kaiser2008}, we have now ways of determining height from
stereoscopic information provided by views from different
angles \citep{Gissot2008,Liewer2009, Gosain2009}.  The three
dimensional reconstruction methods have been developed based on
triangulation technique
\citep{Feng2007,Feng2009,Asch2008,Asch2009,Rodriguez2009}.
These methods use  so called ``tie-pointing" technique, where
same feature is manually located in both images to reconstruct
the 3D coordinates of the feature \citep{Thompson2006}. The
technique uses ``epipolar constraint" to reduce 2D problem to
1D \citep{Inhester2006}. However, for very wide separation
angles these methods present difficulty in identifying common
features.
  Nevertheless, the stereoscopic techniques can still be used to estimate the filament height and inclination even when separation angle is
 large \citep{Gosain2009}.

Observationally it is known that a filament sheet is a suspended structure in the corona with its feet touching the chromosphere periodically. Here, we
present a method for estimating the width and inclination of the filament sheet from its projected width seen in He II 304 \AA\ filtergrams. The width
of the filament sheet and inclination are determined using simple geometric relations under simplifying assumptions regarding the filament sheet. For
 an erupting filament this method gives width-time profile, giving the expansion speed of the filament sheet or flux rope. Further, the full width of
  the filament sheet, measured from its feet to its apex, together with inclination gives an estimate of the filament height above the chromosphere.
  Determining filament inclination prior to their eruptions could be useful to predict the direction in which the material could be ejected, which in
  some cases is highly oblique as found by \cite{Bemporad2009}  by applying triangulation method on an erupting limb prominence.

The present method supplements the triangulation method and the two methods can be used together to cross-check the results for consistency.
 We demonstrate this by applying the two methods to a filament observed by the two STEREO satellites, separated 52.4$^\circ$ apart, and get consistent
 results. The method presented here is best applied when the STEREO separation angle is large because the apparent filament width seen by the two
 satellites is quite different.

In section 2 we describe the method and define the various
angles and notations with the help of an illustration. In
section 3 we apply our method to the STEREO  observations of a
filament eruption during 22 May 2008. Finally, in section 4 we
discuss and conclude the results.

\begin{figure}    %%%%%%%%%%%%%%%%%% FIGURE 1
\centerline{\includegraphics[width=.6\textwidth,clip=]{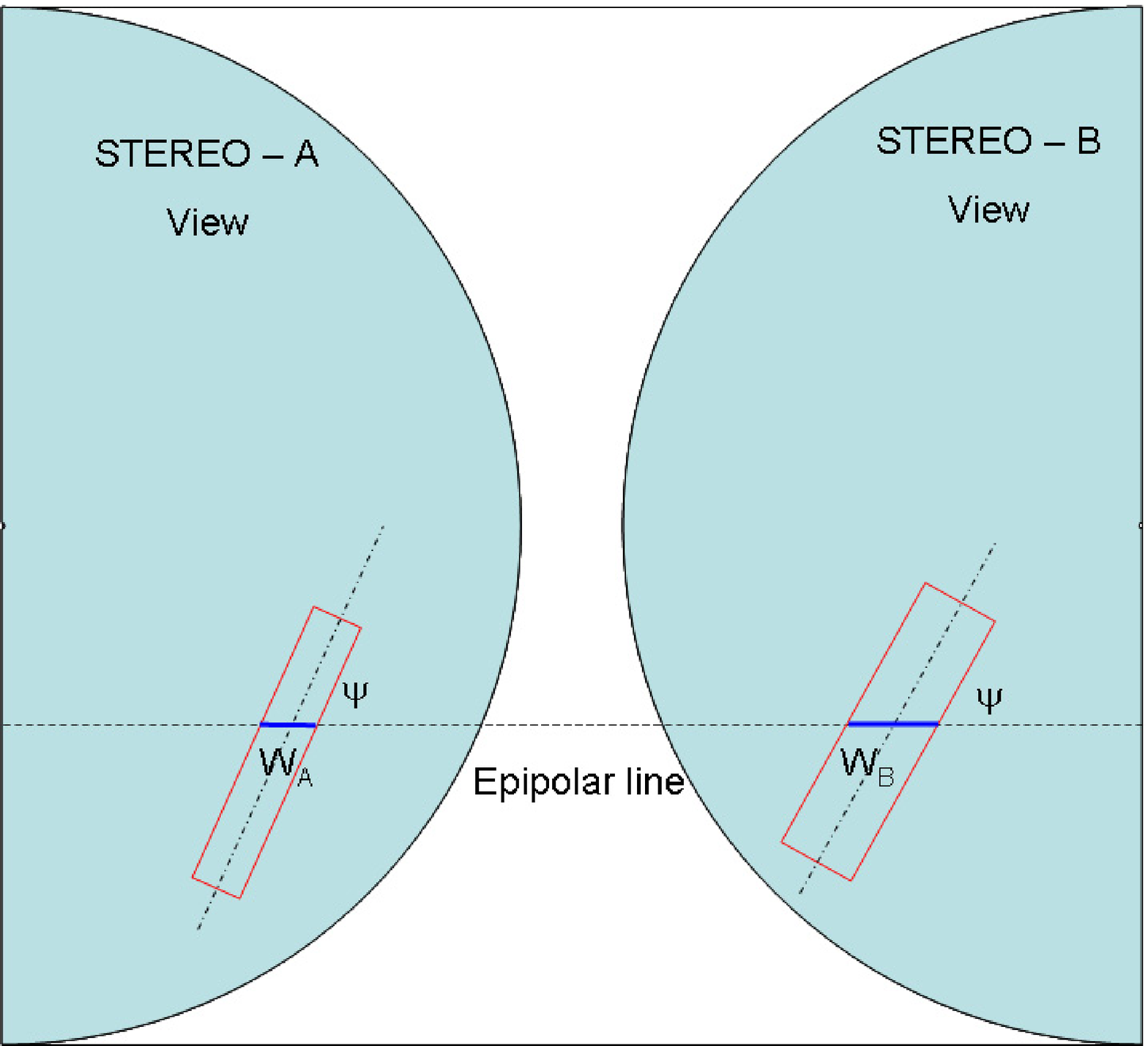}}
\centerline{\includegraphics[width=.8\textwidth,clip=]{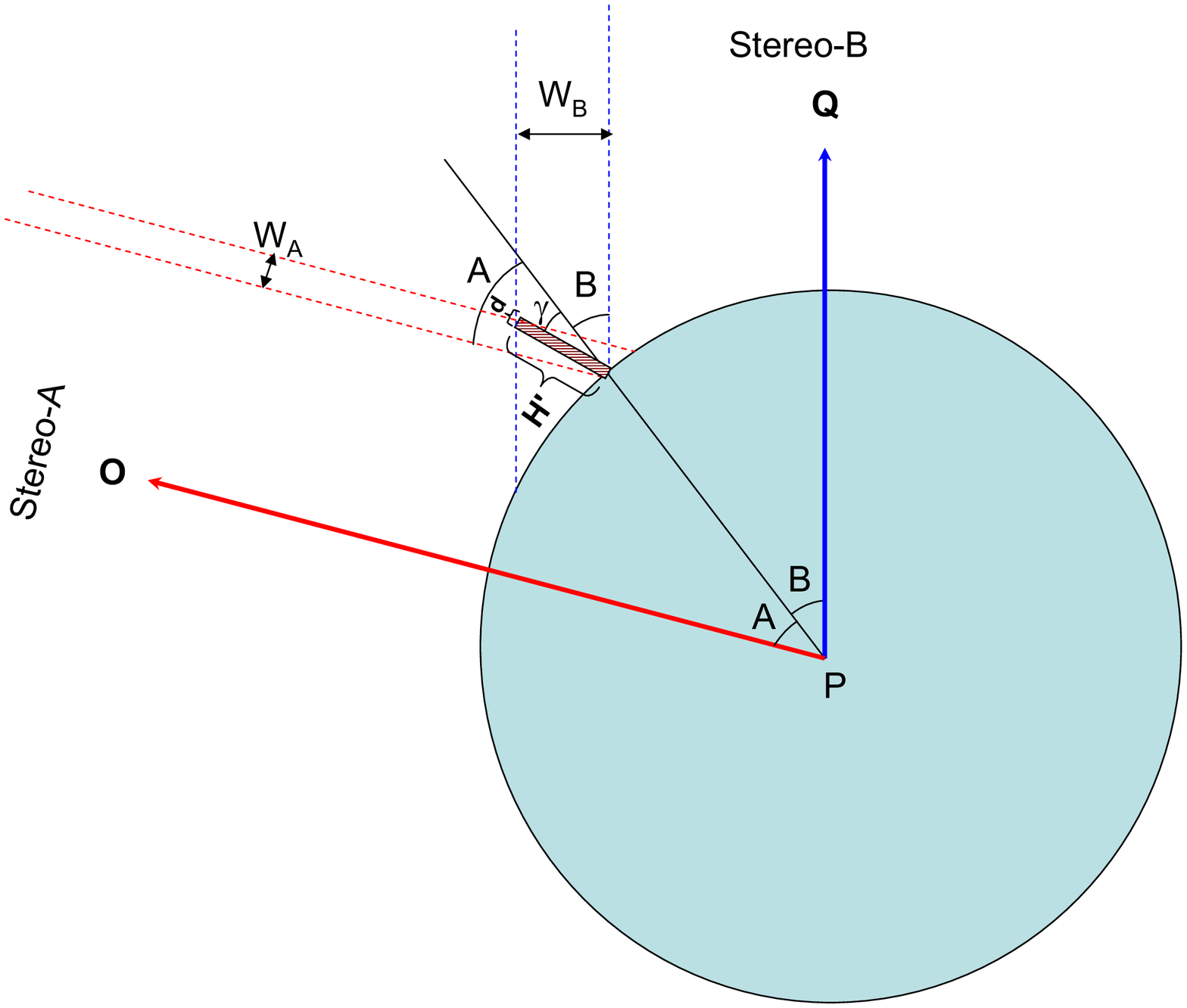}}
\caption{   Top panel (Epipolar plane view):  The two views of the sun from STEREO-A,B are illustrated with filament-axis at an arbitrary
orientation $\Psi$ to epipolar plane. The projected width of the filament measured along epipolar line is $W_A$ and $W_B$.
Bottom panel (Cross-sectional view) :   The blue circle represents cross-section of the sun along a vertical plane through
 epipolar line drawn in top panel. The Sun, STEREO A, B satellites along with a cross-section of the filament along $W_A$ and $W_B$ are shown.
 The segment $H'$ represents the  intersection of filament sheet and epipolar plane. Since filament is oriented at an angle $\Psi$ the width
 of the filament sheet is determined by $H=H' sin \Psi$. For $\Psi=90^\circ$ we will have $H=H'$. The filament sheet is inclined at an angle
 $\gamma$ to the local solar vertical. The angles A and B are heliocentric angles of the filament base in the reference frames of STEREO-A,B
 respectively. P is the central point of this solar cross-section.}
\label{fig:illustrangles}
\end{figure}

\section{Estimation of filament inclination and {\bf width}}
 We now present a technique for estimating the  width and inclination of a filament sheet using  the projected width of the
  filament as observed in He II 304 \AA\ filtergrams observed by the STEREO satellites.  A geometric illustration of this method
  is given in figure~\ref{fig:illustrangles}. The top panel of the figure~\ref{fig:illustrangles} illustrates two views of a filament
  on the solar disk as seen from STEREO-A, B. The filament-axis has an arbitrary orientation $\Psi$ with respect to the epipolar line.
  The projected width of the filament measured along the epipolar line is $W_A$ and $W_B$.  In the bottom panel of figure~\ref{fig:illustrangles},
   the cross-section of the sun taken along a vertical plane passing through epipolar line (drawn in top panel) is shown. The intersection of this
   plane with the filament sheet  can be represented  by the red-hatched segment $H'$. This segment projects widths, $W_A$ and $W_B$ towards STEREO-A
   and B respectively. The relation between $H'$  and the  width of the filament sheet  is $H=H'Sin\Psi$. Further, A and B are the angles
   subtended at point P by the filament base and central meridian longitude in the reference frames of STEREO-A, B respectively. The point P
    is the intersection of epipolar plane and the axis of symmetry (epipolar north-south axis).  The projected width of the filament as seen
     by STEREO A and B is then $W_A=H'sin(A-\gamma) + d ~cos(A-\gamma)$  and $W_B=H'sin(B+\gamma) + d ~cos(B+\gamma)$, respectively. However,
     we can neglect the second term under the simplifying assumptions that, (i) the filament thickness $d$ is much smaller than $H'$ i.e., ($d \ll H'$),
     (ii) the STEREO separation angles $A$ and $B$ are large, and (iii) angle $A$ is not equal to $\gamma$, in which case $W_A$ would measure the
      thickness $d$ of the filament sheet. With these assumptions holding, we can attempt to measure filament inclination and  width as follows:

\begin{equation}
\hspace{1 in} W_A=H' sin(A-\gamma) = H'[sinA~cos\gamma - cosA~sin\gamma]
\end{equation}
%~~~~~~~~~~~~~~~~~~~~~~~ &=&H'[sinA~cos\gamma - cosA~sin\gamma]
\begin{equation}
\hspace{1 in} W_B=H' sin(B+\gamma) = H'[sinB~cos\gamma + cosB~sin\gamma]
\end{equation}

%\begin{eqnarray}

%~~~~~~~~~~~~~~~~~~~~~~~ &=&H'[sinB~cos\gamma + cosB~sin\gamma]
%\end{eqnarray}

since angles A, B and widths $W_{A}$ and $W_{B}$ are known from observations,
we get using Eqns (1) and (2) above \\

Case 1.  $\gamma < A$
\begin{equation}
\hspace{1 in} tan (\gamma)=\frac{C_{1}A_{1}-C_{2}A_{2}}{C_{1}B_{2} - C_{2}B_{1}},\\
\end{equation}

Case 2. $\gamma > A$
\begin{equation}
\hspace{1 in} tan (\gamma)=\frac{C_{1}A_{2}-C_{2}A_{1}}{C_{1}B_{1} - C_{2}B_{2}}
\end{equation}

%\begin{eqnarray}
%~~~~~~~~~~~~~~~~~~~~ C_{1} &=& A_{2}H' cos\gamma -B_{1}H' sin\gamma \\
%~~~~~~~~~~~~~~~~~~~~ C_{2} &=& A_{2}H' cos\gamma -B_{2}H' sin\gamma
%\end{eqnarray}
where,
\begin{eqnarray}
~~~~~~~~~~~~~~~~\hspace{1 in} A_{1} &=& sinA+sinB,\\
~~~~~~~~~~~~~~~~\hspace{1 in} B_{1} &=& cosA+cosB,\\
~~~~~~~~~~~~~~~~\hspace{1 in} A_{2} &=& sinA-sinB,\\
~~~~~~~~~~~~~~~~\hspace{1 in} B_{2} &=& cosA-cosB,\\
~~~~~~~~~~~~~~~~\hspace{1 in} C_{1} &=& W_A -W_B , \\
~~~~~~~~~~~~~~~~\hspace{1 in} C_{2} &=& W_A + W_B
\end{eqnarray}

In order to distinguish between the two cases 1 and 2 above we can use the observation itself. In case 2, when $\gamma > A$  the filament base
 and surface features close-by will be seen on the same side of the spine in both STEREO A and B images, if not then case 1 holds. This is
 illustrated with an example in the next section, where the technique is applied on the real STEREO observations.

Thus, we can get estimate of $\gamma$ from eqn. (3) or (4) depending upon the case applicable, and then using eqn. (1) or (2) we can estimate
$ H'$. The width $H$ of the filament sheet is then given by $H=H' sin\Psi$.

\section{Estimation of filament height}
Let us assume that the filament is not inclined, then the height of the filament sheet is nothing but its width from its feet to its apex.
For estimating the height of the filament we must, therefore,  first locate the feet of the filament in the STEREO images. Then using the
projected width $W_A$ and $W_B$ of the filament from its feet to the spine we can estimate the width $H$ and inclination $\gamma$ of the
 filament sheet using eqn. (1)-(4). We call this width as the full width $H_{f}$ (from feet to apex) of the filament sheet. The height of
 the filament is then given by $H_{f} cos \gamma$. This is demonstrated in the next section.

\begin{figure}[!h]    %%%%%%%%%%%%%%%%%% FIGURE 1
\centerline{\includegraphics[width=0.9\textwidth,clip=]{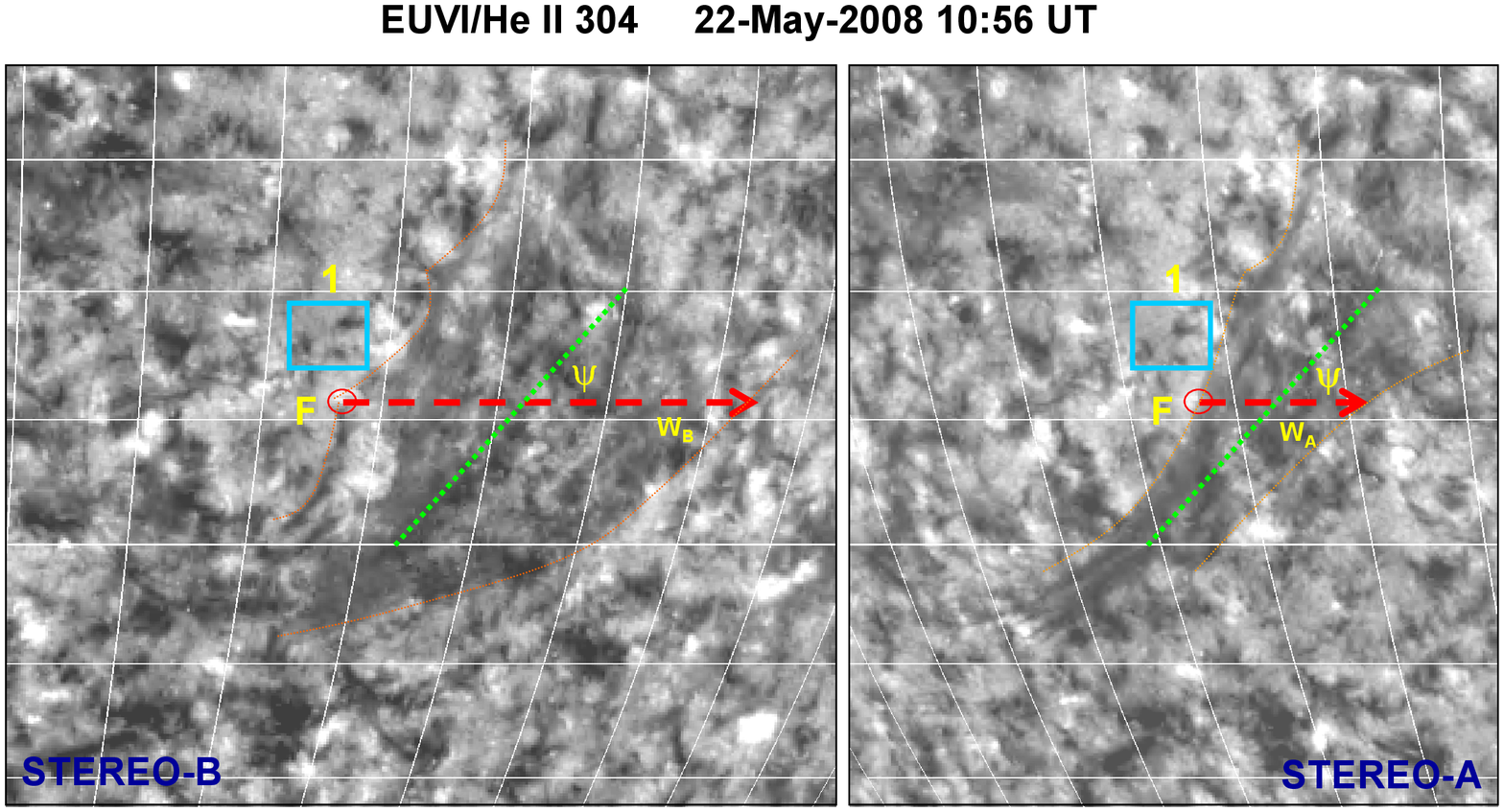}}
\caption{  The He II 304 \AA\ observations of a filament on 22-May-2008 during its disappearance phase at 10:56 UT.
The separation angle of about 52.4 degrees between STEREO satellites shows quite different view of the filament. The spherical
 grid overlaid on the images help to distinguish between surface and elevated features. The location 'F' is identified as the
  feet of the filament. The projected width $W_A$ and $W_B$ is marked by red arrow. Green line marks the orientation $\psi$
  of the filament. Thin orange lines  are drawn to demarcate the boundary of the filament. The features inside box '1' are surface
  features which are seen on left-side of the filament in both images, suggesting that the filament inclination is larger than 'A'. }
\label{fig:illustrdata}
\end{figure}

\begin{table}
\caption{Observed width and position parameters the filament.}
\begin{tabular}{lcccr}
\hline
Parameter && & &  Value  \\
\hline
$W_A$ (Mm)& &~~~~~~~~~~~~~~~~~~~~~~~~~~~~~~~~~~~~~~~~~~~~~~~~~~~~~~~~~~ & & 79\\
$W_B$ (Mm)&& & &  194\\
$\angle A$ (Degrees)& & & & 31\\
$\angle B$ (Degrees)& & & & 20\\
$\Psi$ (Degrees)& & & & 48\\
%$H'$ (Mm) & & & & 200\\
\hline
\end{tabular}
\label{tab:1}
\end{table}

\begin{table}
\caption{Estimated width and inclination of the filament.}
\label{tab:2}
\begin{tabular}{lcr}
\hline
Parameter & Using apparent widths & Using SCC\_MEASURE  \\
\hline
Width, $H$ (Mm) & 148 & 147 \\
Inclination, $\gamma$ (Degrees)& 54 & 47\\
\hline
\end{tabular}
\end{table}

\section{Demonstration on STEREO observations}
 Here we apply this technique on a large filament observed by STEREO during its disappearing phase on 22 May 2008 at 10:56 UT.
 The STEREO images are centered, co-sized, corrected for satellite orientation  to bring them in epipolar view. Figure 2 shows the
 filament as seen by STEREO-A and B in EUVI He II 304 \AA\ wavelength. The filament is outlined by thin orange line. The location
  marked as 'F' is inferred as feet of the filament from its arch like geometry in STEREO-B image on the left panel. Further, the surface
   features inside box `1' are seen on left-side of the filament in both images. While this is what we expect for STEREO-B image for
   STEREO-A image we expect to see these features on right-side if $\gamma < A$ or beneath it (i.e., blocked by filament) if $\gamma = A$.
   This clearly suggests that the filament is inclined by an angle more than $A$. The horizontal red-arrow shows the segment used for
   determination of filament  width and inclination. This segment is chosen because it extends  all-the-way from the spine of the
   filament to its base i.e., feet marked as `F'.  Thus, choosing this segment we can determine the full width $H_f$ and thus the
    height $H_f cos\gamma$ of the filament. The observed parameters of the filament for both STEREO A and B images are given in Table~\ref{tab:1}.
    The values of inclination angle $\gamma$ and segment $H'$ are estimated to be 54 degrees and 200 Mm respectively. The full
     width, $H_f= H'sin \Psi$ is thus 148 Mm and the inclination is 54 degrees. The height of the filament sheet, $H_f cos\gamma$,
     is then estimated to be 87 Mm.

 These values are found to be in good agreement with the values obtained from 3D reconstructions of the same filament using SCC\_MEASURE
 triangulation procedure as described in \cite{Gosain2009}. The triangulation method SCC\_MEASURE gives inclination angle $\gamma$ to be 47 degrees,
 while  width of the sheet $H$ is found to be about 147 Mm. These values are in good agreement (Table~\ref{tab:2}) considering the simplified
 assumptions made in the present method. A difference of 7 degrees in the values of filament inclination determined from the two methods could be
 because of the finite thickness $d$ of the filament (neglected here), which would lead to excess inclination. In which case, a difference
 $\Delta \gamma$ of 7 degrees would mean thickness $d=H/tan (\Delta \gamma)$, which is about 18 Mm for the present case. The interesting
 possibility of using both the methods simultaneously to infer filament  thickness $d$ along with  width $H$ and inclination $\gamma$
 is deferred to another study, where large number of cases will be tested for consistency. For now, one should be careful in applying the method
 and take into account simplifying assumptions under which estimations are made.

\section{Discussion and Conclusions}
  When the separation angle between the twin STEREO satellites is quite large it becomes difficult to identify common features in the two images.
  Such identification of common features in STEREO images is quite important to compute the 3D coordinates using triangulation techniques
   \citep{Inhester2006,Asch2008}. Also, the other methods like optical flow technique used by \citep{Gissot2008} are difficult to use
   with widely separated angles. In such scenario our method can be used to supplement the estimates of filament  width and inclination
    to cross-check the values obtained by other methods.

The method proposed above has obvious limitations and is applicable under simplified assumptions. These assumptions are:

(i)	A filament is a rectangular sheet of thickness $d$, width
$H$ and length $L$.  Further, the sheet is assumed to be thin
i.e., $d \ll H$. This is generally true for quiescent
prominences which are very tall and thin, especially at the
apex of the filament, which determines the extent of filament
projection.  Application to active region filaments, therefore,
should not be attempted as it may give large errors.

(ii) It is possible to locate the feet of the filament in STEREO images. We demonstrated in the example presented in this paper that it is
 possible to carefully locate filament feet (extending down to the chromosphere). One must carefully select such portions to estimate the
 filament height and inclination. The height of the filament can be estimated when the projected widths $W_A$ and $W_B$ extend from filament
 feet to its apex. In this case the technique mentioned above would estimate inclination $\gamma$ and full width $H_f$ (from its feet to its apex)
 of the filament sheet, giving height as $H_f cos \gamma$.

(iii)	The third assumption is that, it is possible to judge from the images whether the inclination angle gamma is larger, equal or smaller than
angle A (Eq. 3 and 4). We showed using a set of STEREO observations that this is possible for a large prominence by locating position of surface
features with respect to filament base. However, such determinations could be difficult in case of active region filaments which are generally
low-lying and do not show wide projections.

Departure from these assumptions could lead to large errors which are not quantified yet. Especially for active region filaments, as argued above,
the errors would be large. In order to quantify such errors we plan to use our technique on simulated 3D structures like extrapolated magnetic field
lines in spherical geometry, in our future work.

\begin{acknowledgements}
  SG acknowledges CEFIPRA funding for his visit to  Observatoire de Paris, Meudon, France under its project No. 3704-1. This work was supported
  by the European network SOLAIRE (MTRN\_CT\_2006\_035484).
\end{acknowledgements}

%\bibliographystyle{copernicus}
%\bibliography{ref}

\end{document}